# Oedometric compression of a granular material: computation of energies involved during breakage with a discrete element modelling

*François* Nader[1#], *Claire* Silvani[1*], and *Irini* Djeran-Maigre[1]

[1] Univ Lyon, INSA Lyon, GEOMAS, F-69621 Villeurbanne, France

**Abstract.** A numerical model able to simulate the grain breakage with the discrete element method, using the "Non-Smooth Contact Dynamics" is presented. The model reproduces 3D grains having complex shapes and is tested in single grain and in oedometric compressions. Numerical simulations are then carried out to evaluate the different energies active during breakage (surface creation and redistribution energies). The surface creation energy is estimated. Results are closed to the ones found in the literature.

## 1 Introduction

Civil engineering structures built using rockfill undergo substantial deformation throughout their entire lifespan, due to the possible breakage of the material and the subsequent rearrangements under their own weight. Grain breakage plays a major role in the behaviour of these coarse granular materials. These breakages become more frequent with the increase in the angularity and size of the grains.

Despite the precautions put in place during construction to limit settlement (compaction, watering, etc.), these structures can experience significant settlements, and even reach dangerous levels in the case of dams. It is therefore important to understand the behaviour of these materials. Since the 1960s, the experiments carried out have shown the sensitive points of these materials [1-4]. Experimental tests prove to be the most reliable method for apprehending the behaviour of these materials under different types of loadings, but they have the disadvantage of being impractical and expensive, especially when it comes to large samples requiring large experimental devices. Numerical modelling then seems to be an alternative to understand and predict the behaviour of rockfill structures, and to propose solutions to avoid the risks that this type of structure can possibly present.

Given the discontinuous nature of these materials, the discrete element method seems to be the best suited method to describe the behaviour of these granular materials. The present work thus proposes a three-dimensional polyhedral (convex) grain model, capable of breaking, used in discrete numerical simulations to realistically reproduce the grain shapes present in nature. This grain model is tested in single grain compression tests and then implemented into a multigrain oedometric sample. A method is then proposed to estimate the different energies involved during the breaking of the grains.

## 2 Breakable grain model and single grain compression test

This proposed grain model (figure 1) considers a grain in the form of an assembly of tetrahedral particles, linked by a cohesive law based on a Mohr-Coulomb failure criterion. The tetrahedral particles are generated using a Delaunay triangulation of the initial volume. A scale parameter allows to control the size and therefore the number of particles. No porosity is considered within this "macro-grain". The "pieces" generated by the breakage end up with similarly angular shapes. This model is implemented in the LMGC90 open-source code platform, based on the Non-Smooth Contact Dynamics method [5], which assumes that all bodies are totally rigid, like the contacts not allowing interpenetration. The Non-Smooth Contact Dynamics method differs from the Smooth Contact methods, also known as Soft Contact Methods. In soft contact approaches, a normal and a tangential stiffness are defined at the contact level, which allows the contact to behave in a manner similar to a spring. These stiffnesses allow an overlap at the contact level, and this overlap is then used to compute the forces at contact when solving the equation of Newton's second law of motion. In Non-Smooth Contact Dynamics, there is no explicit relationships to express the contact force, its calculation is instead implicit and based on shock laws.

The input parameters of the model are the normal ($C_n$) and tangential ($C_t$ with $C_t = \mu C_n$) cohesions, and the coefficient of friction between the particles $\mu$. When the stress state reaches the Mohr-Coulomb failure

* Corresponding author: claire.silvani@insa-lyon.fr
# Current address : Institut de Recherche Technologique RAILENIUM, F-59300, Famars, France

criterion, the cohesion between particles is lost. From then onwards, the particles interact with each other by contact and friction. Diametral compression tests were conducted on a single grain geometry with varying cohesions, and on grains of different sizes (figure 2) [6]. The local cohesive surface is kept constant as the maximum grain size increases, meaning that the number of subgrains (tetrahedral particles) increases with the grain size.

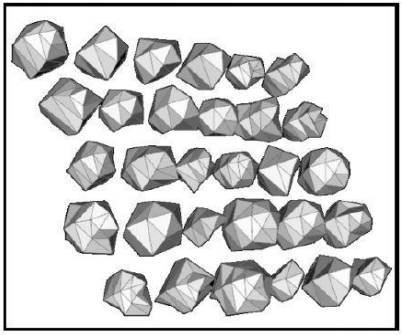

**Fig. 1.** Sample of randomly generated breakable grains.

The breakage stress is computed according to the definition proposed by Jaeger [7]:

$$\sigma_R = F_R / d^2, \quad (1)$$

where $F_R$ is the breakage force, and $d$ the maximum grain dimension. The scale effect could be simulated by the model (the breakage stress decreases as the grain size increases and the probability of finding a defect in the volume increases), according to Weibull's probability of failure theory [8]. Additional tests to evaluate the sensibility of the grain's mesh (number of subgrains) were studied and showed that the breakage force is proportional to the size of the subgrains' cohesive surface as expected [6].

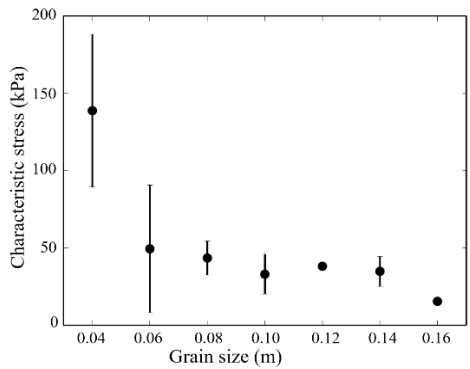

**Fig. 2.** Compression test of a grain and breakage stress $\sigma_R$ for grains of different sizes for a set of parameters ($C_n = 10^3$ kPa, $\mu = 0.4$).

# 3 Estimation of the various energies involved during oedometric compression

## 3.1 Oedometric tests on breakable and unbreakable grains

A sample of 850 grains is generated with an initial uniform grain size distribution and a grain size of 4 cm (figure 3). The size of the sample before compression is 30 x 30 x 26 cm. Two types of simulations are carried out, with breakable and unbreakable grains. The breakable grains sample includes ~ 12 particles per grain (or 10,700 tetrahedra). The curves of the two materials (figure 4) show an expected result for an oedometric compression test.

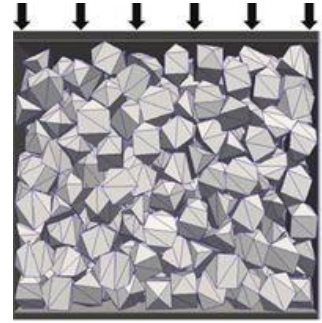

**Fig. 3.** Numerical sample of grains.

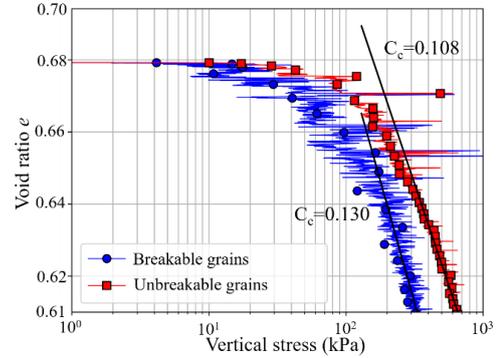

**Fig. 4.** Oedometric e-logσ curves (median curves) for breakable and unbreakable grain samples.

## 3.2 Definition of energies involved in the breakage phenomenon

First, we calculated the external work as follows:

$$W_{ext} = R_{sup} \times \Delta u, \quad (2)$$

where $R_{sup}$ is the reaction on the superior plate and $\Delta u$ is the displacement. The comparison of the external works (figure 5) shows that during compression (from ε = 4%), the sample of unbreakable grains stiffens, and the external work of the sample of breakable grains becomes lower, because of the grains' greater freedom of movement.

A breakage energy is then computed, by considering it equal to the difference between the two curves in figure 4. This breakage energy corresponds physically to the sum of two energies: an energy necessary for the creation of new surfaces by breakage of grains $E_{surf}$, and an energy due to the rearrangement of fragments resulting from the fracture $E_{redist}$ [9, 10] proposed by equation (3) in order to compute the surface creation energy *per unit of volume* produced by the fracture of the grains :

$$\Delta E_{surf_{volume}} = \Gamma \frac{\Delta S}{V_S(1+e)}, \quad (3)$$

where $\Gamma$ (N/m) is the specific area energy based on the theory of Griffith [11], $\Delta S$: the newly created surface, $V_S$ : the solid volume, $e$ : void ratio, therefore $V_S(1+e)$ is equal to the total volume.

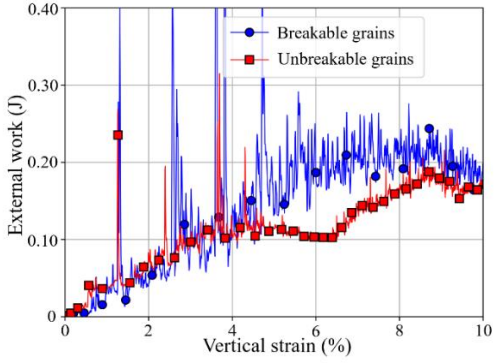

**Fig. 5**. Evolution of the external work during the oedometric compression of samples of breakable and unbreakable grains.

The total plastic work increment $\Delta W_{plast}$ generated during loading is equal to the sum of the energy dissipated by friction $\Delta E_{diss}$ and the energy due to the breakage of the grains $\Delta E_{rupt}$. Taking into account that the fracture energy is in turn divided into a surface creation energy $\Delta E_{surf}$ and a redistribution of kinetic energy via the broken fragments $\Delta E_{redist}$, and using the ratio $R = \frac{\Delta E_{redist}}{\Delta E_{surf}}$ defined by Russell [12], we obtain the equation (4) :

$$\Delta W_{plas} = \Delta E_{diss} + \Delta E_{rupt} = \Delta E_{diss} + \Delta E_{surf}(1+R). \quad (4)$$

Using equations (3) and (4), and considering the total surface creation energy (and not per unit volume), $\Delta E_{rupt}$ can be written as follows:

$$\Delta E_{rupt} = \Delta S\ \Gamma(1+R). \quad (5)$$

The next step consists of estimating the specific energy value $\Gamma$.

### 3.3 Estimation of the specific surface energy $\Gamma$

The specific energy $\Gamma$ is an intrinsic parameter specific to the material subjected to breakage. In order to estimate the value of $\Gamma$ specific to the numerically modelled material used in the oedometric simulations, a series of simulations is performed by dividing a single grain into tetrahedral particles, while keeping the rest of the grains unbreakable, and the sample is completely identical initially (figure 6). Eleven simulations are carried out, by selecting a random grain for each case. The proposed method assumes that $\Delta E_{rupt} = \Delta E_{surf}$, so that all the fracture energy is transformed into surface creation energy. In fact, given that a single grain breaks, the part of $\Delta E_{rupt}$ which comes from the redistribution $\Delta E_{redist}$ can be neglected (i.e. $R = 0$). The deviation on the curves between the external work of the single breakable grain and non-breakable samples allows the fracture energy to be computed using the same principle explained previously.

For each simulation, the values of $\Delta E_{rupt}$ and $\Delta S$ are recorded. Given the large value of the standard deviation from the computed mean value, and the presence of points far apart from the mean, the normal statistical distribution is used. We thus obtain the following average value for $\Gamma = 10$ N/m. This estimated value of the specific surface energy is of the same order of magnitude as the values proposed in the literature [13, 14].

### 3.4 Evolution of redistribution and surface creation energies

In this part, the oedometric simulation of the sample with 850 breakable grains is repeated. We try to evaluate the part of the surface and redistribution energies in the simulation where the grains can break. The value of $\Gamma$ computed in the previous paragraph is used to calculate the redistribution energy as follows:

$$\Delta E_{redis} = \Delta E_{rupt} - \Delta E_{surf} = \Delta E_{rupt} - \Gamma \times \Delta S. \quad (6)$$

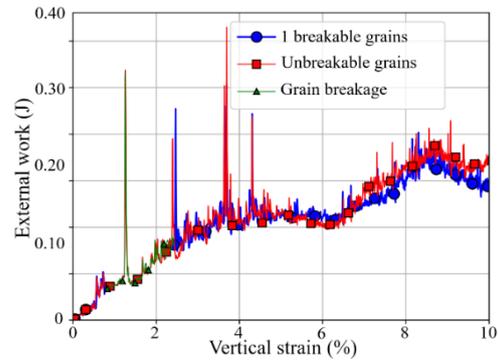

**Fig. 6**. Example of an oedometric compression of a single breakable grain among a set of non-breakable grains: evolution of external work.

Figure 7 shows the evolution of the redistribution and surface creation energies as a function of the vertical deformation for $\Gamma = 10$ N/m. The two energies decrease with respect to the deformation, but the slopes of the regression lines show a greater decrease in $\Delta E_{surf}$ compared to $\Delta E_{redist}$. These results therefore indicate that at the onset of compression, the dissipation of fracture energy in the form of surface creation energy is greater than the dissipation in the form of redistribution energy. But as the compression increases, the order seems to change, the dissipation of energy by redistribution seems to become greater. Additional simulations are required to validate this tendency, but these primary results are confirmed by the work of Ovalle et al. [4], as explained hereafter.

To validate these results, comparisons are made with the experimental study by Ovalle et al. [4], performed on sand. The experimental results have shown that $\Gamma(1+R)$ is low for low values of $\Delta E_{rupt}$ therefore under low stresses and increases for high values of $\Delta E_{rupt}$ (under high stresses).

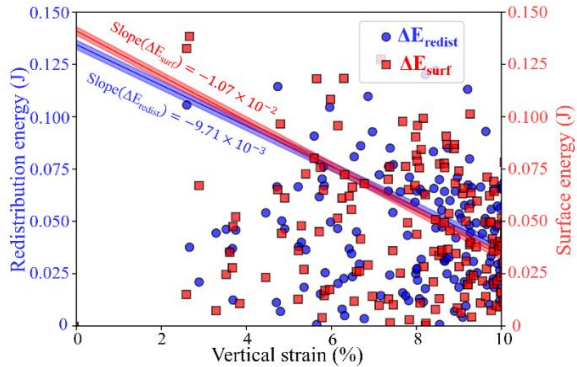

**Fig. 7**. Evolution of redistribution $\Delta E_{redist}$ (left axis) and surface $\Delta E_{surf}$ (right axis) energies as a function of deformation during the oedometric compression of 850 breakable grains. The shaded area around the regression lines shows the standard deviation.

This indicates that under low stresses, the surface creation energy $\Delta E_{surf}$ is greater than the redistribution energy $\Delta E_{redist}$, but with the advancement of the loading, this order is reversed. From a physical point of view, Ovalle et al. [4] offered the following explanations:
- under low stress, the contact forces acting on the breaking grains can be transmitted to neighbouring grains without much rearrangement.
- under high stresses, neighbouring grains cannot withstand the transmitted forces, and a global rearrangement of the grains is necessary to withstand the loading.

By comparing with the numerical results obtained for a value of $\Gamma = 10$ N/m, the same tendency is observed: a higher slope (in absolute value) for the energy of creation of surface $\Delta E_{surf}$ indicates that this energy decreases in a more important way, therefore the value of $R=(\Delta E_{redist})/(\Delta E_{surf})$ increases, and consequently the value of $\Gamma(1+R)$ increases.

## 4 Conclusion

A numerical breakable grain model is developed to study the energies involved in the breaking of grains. The model is validated for single grain compression tests, then the grain model is implemented into a multigrain sample with oedometric loading.

First, the fracture energy $\Delta E_{rupt}$ is computed as the difference between the external work of the breakable-grain and non-breakable oedometric simulations. Monitoring this breakage energy makes it possible to observe phenomena of blockage of grain movement followed by grain breakage.

Then several simulations, where a single grain is breakable, are used to estimate a value of the specific surface energy $\Gamma = 10$ N/m. This value is then used to calculate the surface creation $\Delta E_{surf}$ and kinetic energy redistribution energies due to fracture $\Delta E_{redist}$ in the sample where all the grains are breakable. The qualitative comparison of the variation of these energies during compression is in agreement with the results obtained experimentally by [4]. Under low compression, the dissipation of fracture energy in the form of surface creation energy is greater than that in the form of redistribution of kinetic energy. As compression advances, this order seems to change, the dissipation of energy by redistribution seems to become greater. Additional simulations are required to validate this tendency.

Further investigations will have to be carried out to verify also the evolution of the value of the specific surface energy $\Gamma$ depending on the type of failure (mode I or II). Additional tests to evaluate the sensibility of the internal configuration of breakable grains (number and geometry of tetrahedral particles) in the oedometric response will also provide interesting insight on the breakage model.